\documentclass{elsart3p}
\usepackage{amsmath}
\usepackage{amssymb}
\usepackage{dcolumn} 
\usepackage{bm} 
\usepackage{color}
\usepackage{ifpdf}
\ifpdf
    \usepackage[pdftex]{graphicx}
    \usepackage{epstopdf}

\else
    \usepackage{graphicx}

\fi
\usepackage{longtable}
\usepackage{hyperref}
\usepackage{subfigure}
\usepackage{multirow}

\journal{Nuclear Instruments and Methods A}
\begin{document}
\begin{frontmatter}

\title{Validation of spallation neutron production and propagation within Geant4}

\author[UW]{M.G.~Marino\corauthref{cor}} \ead{mgmarino@u.washington.edu}, \author[UW]{J.A. Detwiler}, \author[LBNL]{R. Henning\thanksref{rhenningAdd}}, \author[UW]{R.A. Johnson}, \author[UW]{A.G. Schubert}, \author[UW]{J.F. Wilkerson}
\address[UW]{
Physics Department, University of Washington, Seattle, Washington
}
\address[LBNL]{
Lawrence Berkeley National Laboratory, Berkeley, California
}
\corauth[cor]{Corresponding author.}
\thanks[rhenningAdd]{Currently at: University of North Carolina, Chapel Hill, NC}

\begin{abstract}
Using simulations to understand backgrounds from muon-induced neutrons is important in designing next-generation low-background underground experiments.
Validation of relevant physics within the Geant4 simulation package has been completed by comparing to data from two recent experiments.  Verification focused on the
production and propagation of neutrons at energies important to underground experiments.  Discrepancies were observed between experimental data and the
simulation.  Techniques were explored to correct for these discrepancies.  
\end{abstract}
\end{frontmatter}

\section{Introduction}

In ultra-low background experiments such as the one proposed by the \textsc{Majorana} collaboration~\cite{Aalseth:2004yt}, it is critical to develop a well-understood
background model within a verified simulation framework.  Accomplishing this involves developing a simulation package which can sufficiently simulate physics
processes and detector response to generate an accurate estimation of background events for a given detector.  To this end, the \textsc{Majorana}
collaboration has joined with the GERDA collaboration~\cite{GERDA} to design a simulation and analysis software package (MaGe~\cite{Bauer:2006rr}) based upon the
CERN software Geant4~\cite{Geant4,Allison:2006ve} and ROOT~\cite{ROOT}.  

Cosmogenic backgrounds can be significantly reduced by performing experiments deep underground.   However, even at large depths cosmic-ray muons can generate untagged neutrons over a wide energy range (up to $\sim$1~GeV) which may result in significant backgrounds.  
Simulations of fast neutrons are difficult due to the uncertainties in and disagreement between theoretical models.  
The fact that primary spallation neutrons may themselves produce secondary neutrons and hadronic showers further complicates calculations and makes benchmarking these processes difficult.
Understanding the background created by such neutrons involves the simulation of their creation, propagation and interaction with the detector, as well as the
verification of all associated code.  Previous works have analyzed and compared simulations with experimental data, mainly using the FLUKA~\cite{Ferrari:2005zk}
simulation package (see~\cite{Wang:2001fq,Kudryavtsev:2003au}).  A comparison between data and the FLUKA and Geant4 simulation packages has been completed as
well~\cite{Araujo:2004rv}.  Studies estimating the background from muon-induced neutrons have been performed for the CRESST experiment~\cite{Wulandari:2004bj} as
well as more generally for underground experiments~\cite{MEI06}.  

This work builds on previous studies (especially~\cite{Araujo:2004rv}) to verify neutron creation and propagation and quantify related systematic uncertainties
within the Geant4 simulation package.  New simulations have been performed of two relevant experiments: one at CERN~\cite{Chazal:2001sm} that used a 190~GeV muon beam incident upon a variety of materials; and another at SLAC~\cite{SLAC1} that employed a 28.7~GeV
electron beam incident upon an aluminum beam dump.
The simulation of the CERN experiment
is focused on validating neutron production from muons and the simulation of the SLAC experiment is aimed primarily at testing the propagation of neutrons and associated showers.

Geant4 allows a user-customizable physics list that may be changed to suit the requirements of the simulation.  The physics list used for the simulations was essentially the ``QGSP\_BIC\_HP" list.
The relevant hadronic physics utilized the Binary Cascade for nucleons (below 10~GeV) and parameterized high- and low-energy models to describe pions.  
 At higher hadron energies ($>$12~GeV), a quark-gluon string (QGS) model was employed to model inelastic interactions.  High-energy and low-energy parameterized models bridged the energy regions not covered by the Binary Cascade and QGS models.  Both the Binary Cascade and QGS models employ a pre-compound model to describe the nuclear system while it is not in a state of equilibrium.  At lower energies (0-20~MeV), neutron inelastic interactions were handled by a high-precision, data-driven model.  The G4MuNuclearInteraction was used to model direct interaction between muons and nuclei.  These interactions, also referred to as ``muon spallation" interactions, involve an exchange of virtual photons between the muon and the nucleus and are a significant source of theoretical uncertainty~\cite{Wang:2001fq}.  Geant4 handles the virtual photons as effective pi+ and pi- particles which interact with the nucleus using parameterized models.  For each simulation, several versions of Geant4 from 7.0 onwards were used during development, but the final results presented here are from versions 8.1 (SLAC) and 8.0 (CERN).  For more information on the various models discussed here, please see the Geant4 online documentation~\cite{Geant4}.

\section{CERN NA55 Experiment}\label{sec:CERN}

The NA55 experiment at CERN measured the double-differential cross section for neutrons emitted at 45, 90 and 135 degrees from a 190~GeV muon beam incident upon three different materials: graphite, Cu, and Pb.  A complete description of the experiment can be found in~\cite{Chazal:2001sm}.  

\subsection{Simulation}

The CERN NA55 experiment has been simulated by Ara\'{u}jo et al.~using Geant4 and FLUKA and results from these simulations are presented in~\cite{Araujo:2004rv}.  However, that work did not consider a comparison between the measured and simulated double-differential cross sections (i.e.~$d^2\sigma/dE d\Omega$).  A simulation has been completed to compare to this quantity as well as provide updated results using a recent version of the Geant4 package.  This new simulation generated results consistent with those found in~\cite{Araujo:2004rv}.

The simple simulation geometry consisted of a cylindrical beam target placed in the center of a 2.2~m radius sphere filled with air.  The materials and dimensions of the different beam targets are listed in Table~\ref{tab:CERNDim}.  A pencil beam of 190~GeV muons was incident upon the center of the face of the cylindrical beam dump.  For each neutron passing through the bounding sphere, the energy and angle relative to the beam were recorded.  Particles were only propagated within this bounding sphere and killed once they exited the sphere.

\begin{table}
\caption{Dimensions of cylindrical beam targets used in the CERN NA55 experiment.  (See Reference~\cite{Chazal:2001sm}.)}
\label{tab:CERNDim}
\centering
\begin{tabular}{l|m{.2\columnwidth}|m{.2\columnwidth}|m{.2\columnwidth}}
Material & Length [cm] & Radius [cm] & Density [g/cm$^3$] \\
\hline
\hline
Graphite & 75 & 4 & 2.26 \\
Copper & 25 & 5 & 8.99 \\
Lead & 10 & 10 & 11.35 \\
\hline
\end{tabular}
\end{table}

\subsection{Results}

\begin{figure*}[ht]
\centering
	\subfigure[$\theta=$45$^{\circ}$]{\label{fig:Graphite_45Deg} \includegraphics[width=.9\columnwidth]{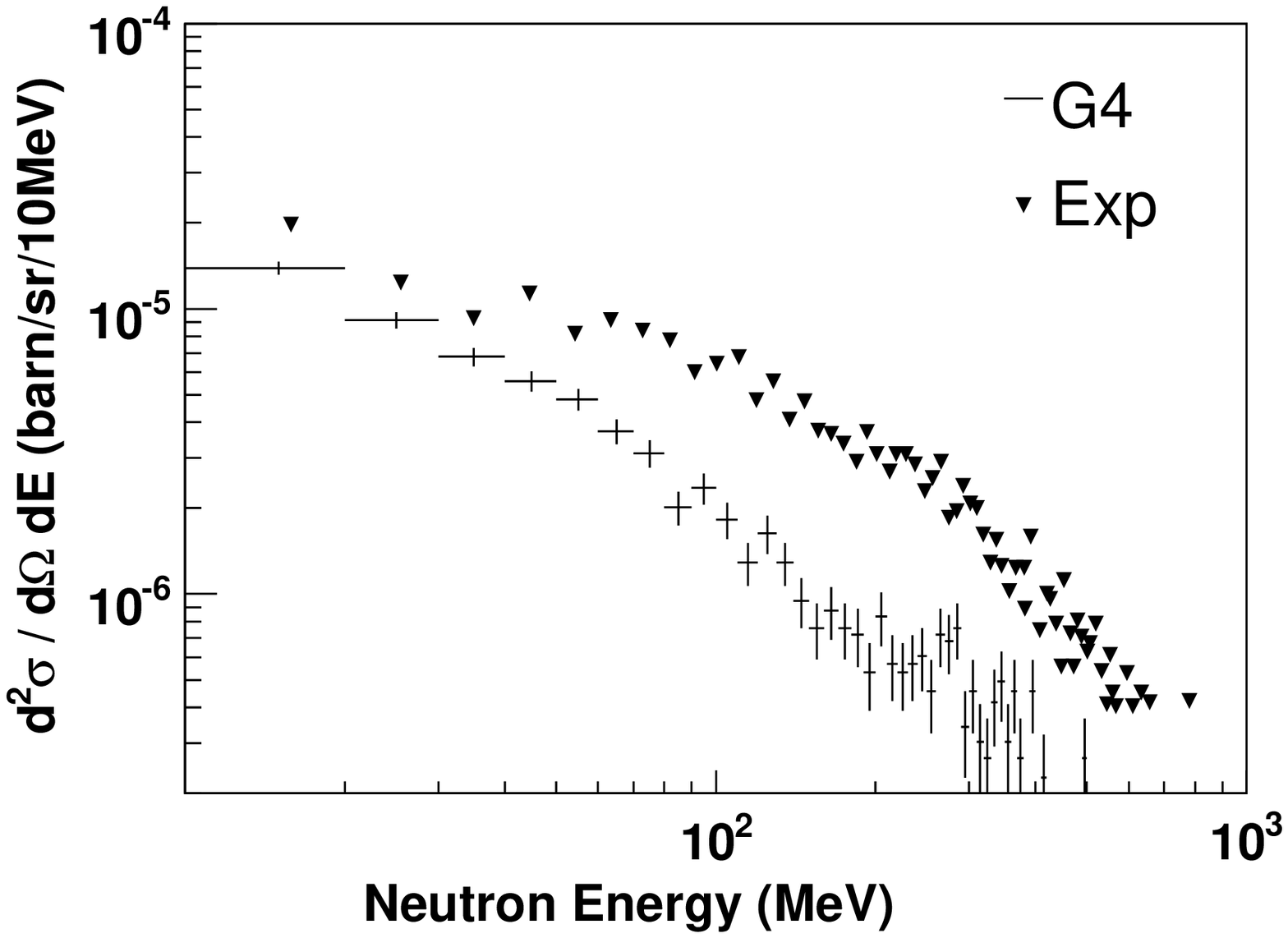}}
	\subfigure[$\theta=$90$^{\circ}$]{\label{fig:Graphite_90Deg} \includegraphics[width=.9\columnwidth]{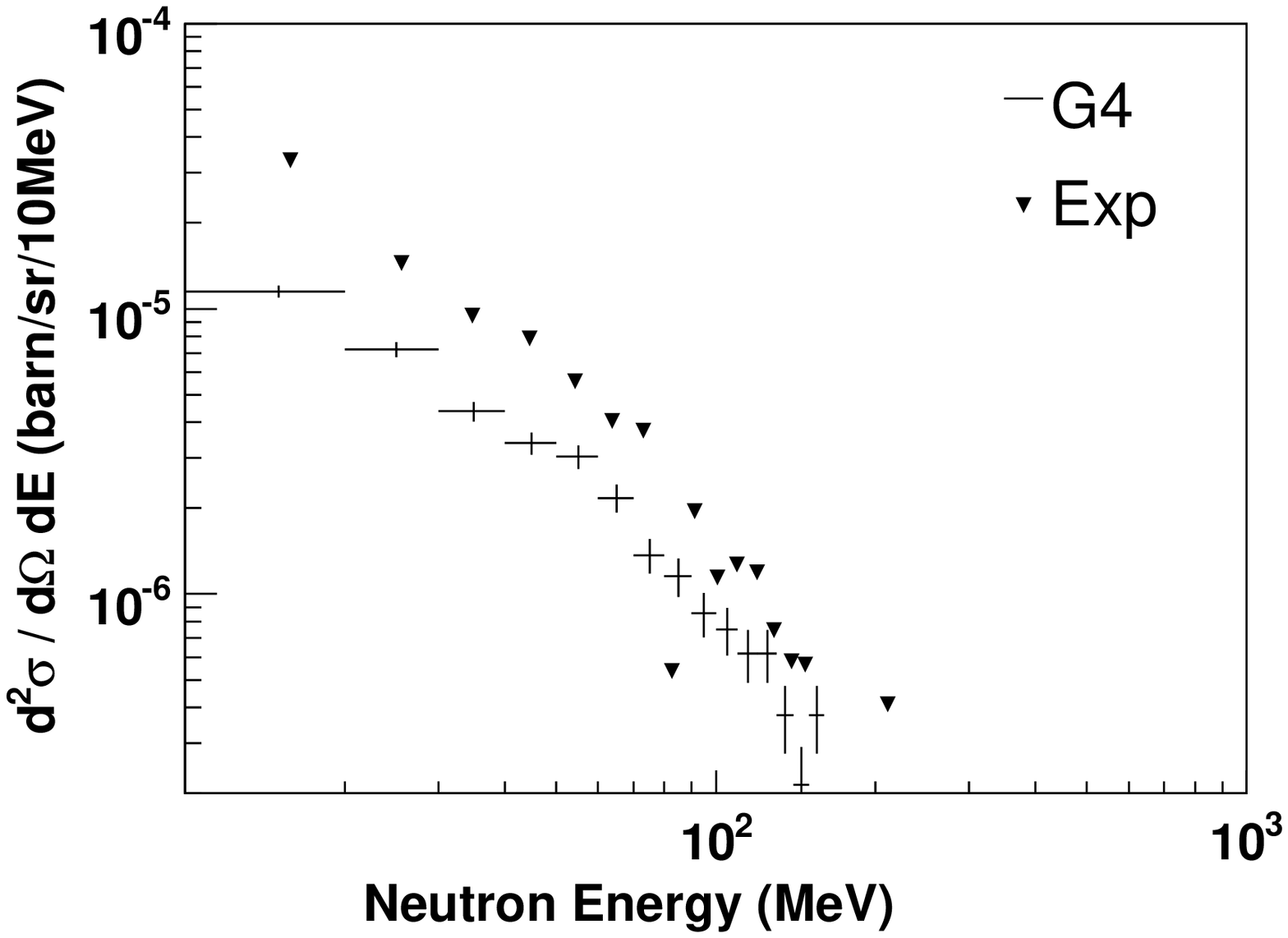}}
	\subfigure[$\theta=$135$^{\circ}$]{\label{fig:Graphite_135Deg} \includegraphics[width=.9\columnwidth]{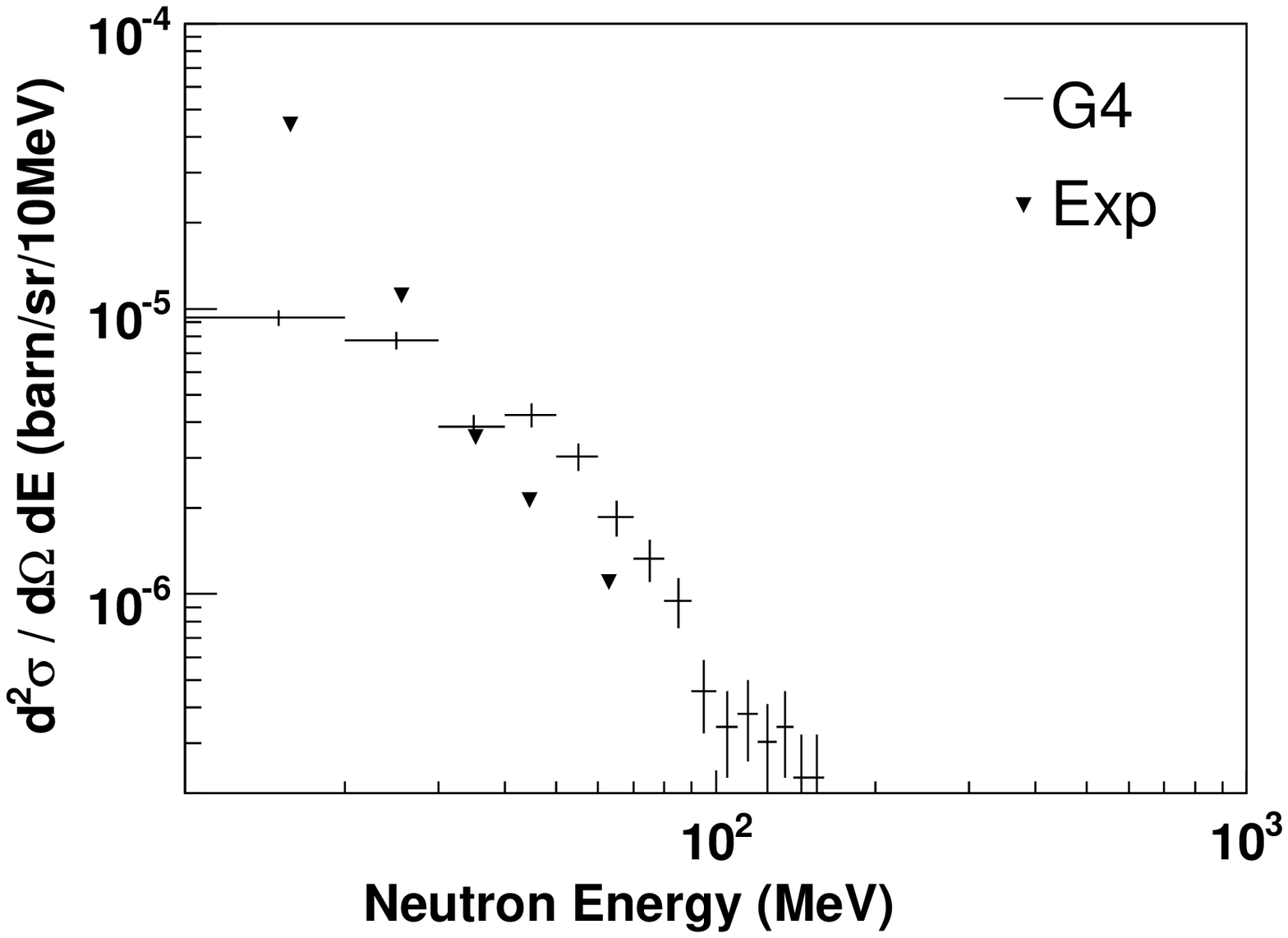}}
	\subfigure[Angular distribution.]{\label{fig:Graphite_CosTheta} \includegraphics[width=.9\columnwidth]{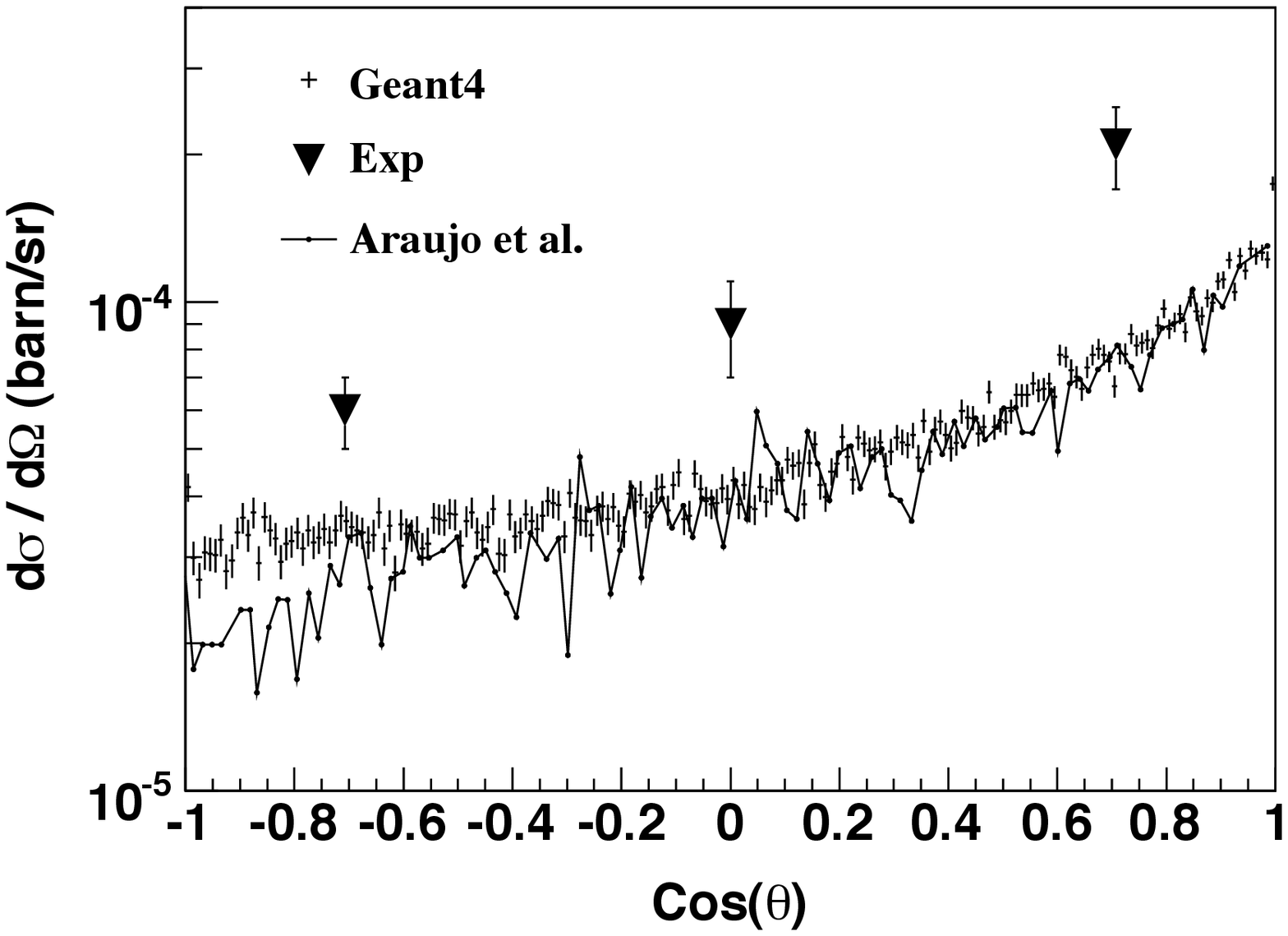}}
\caption{Comparison between the NA55 experiment and simulations for the graphite beam target. 
The angular distribution plot (Figure~\ref{fig:Graphite_CosTheta}) includes data from Ara\'{u}jo et al.'s simulation of CERN NA55~\cite{Araujo:2004rv} (points connected
by lines) demonstrating that this work agrees with the previous simulation with the exception that this work exhibits an increased cross section at back-scattering angles.
The error bars presented on the Geant4 data are statistical; error bars are not included on the experimental energy spectra since these
could not be obtained from a scanned image of the data from Ref.~\cite{Araujo:2004rv}.} 
\label{fig:CERNGraphite}
\end{figure*}

\begin{figure*}[ht]
\centering
	\subfigure[$\theta=$45$^{\circ}$]{\label{fig:Copper_45Deg} \includegraphics[width=.9\columnwidth]{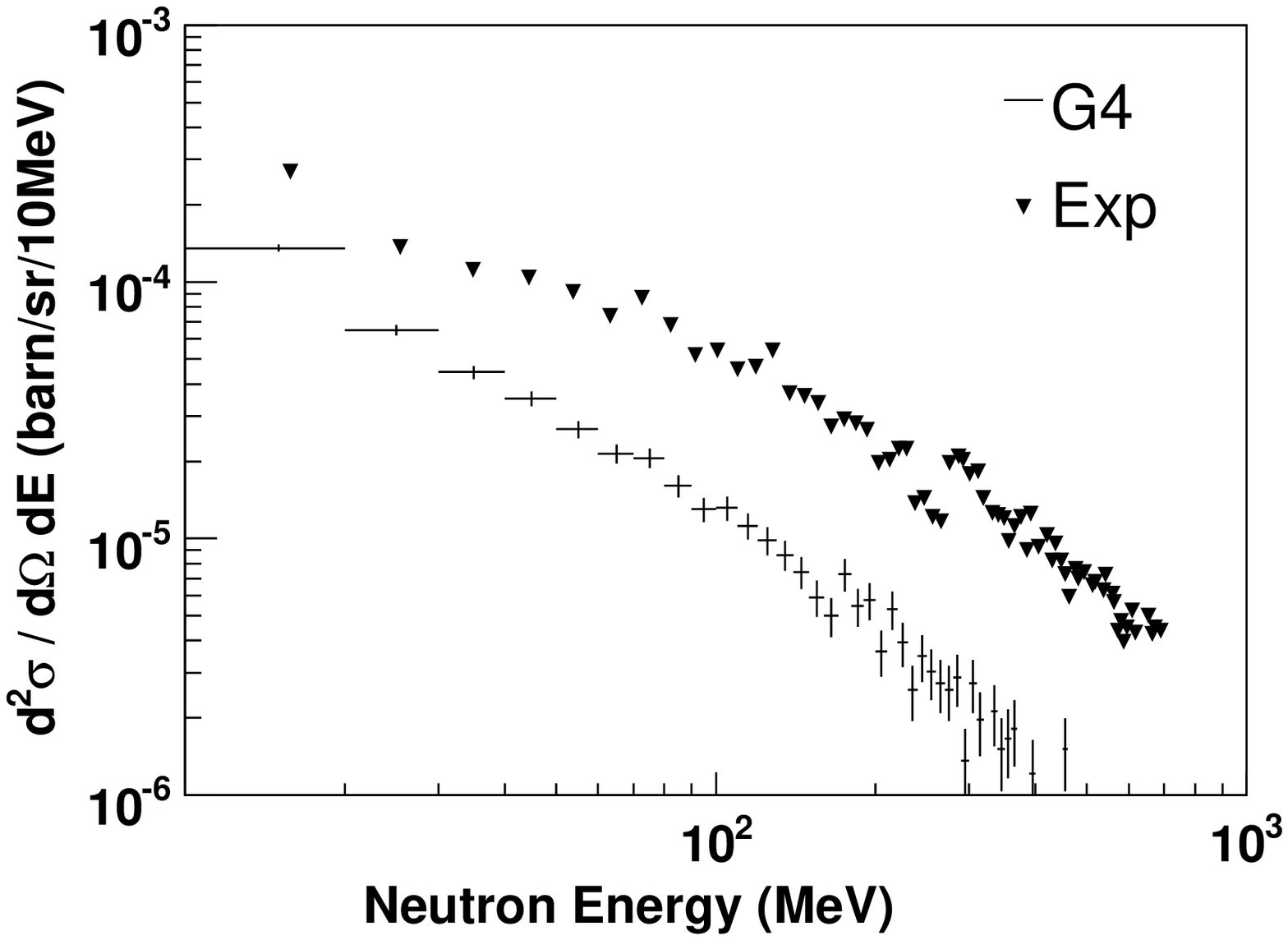}}
	\subfigure[$\theta=$90$^{\circ}$]{\label{fig:Copper_90Deg} \includegraphics[width=.9\columnwidth]{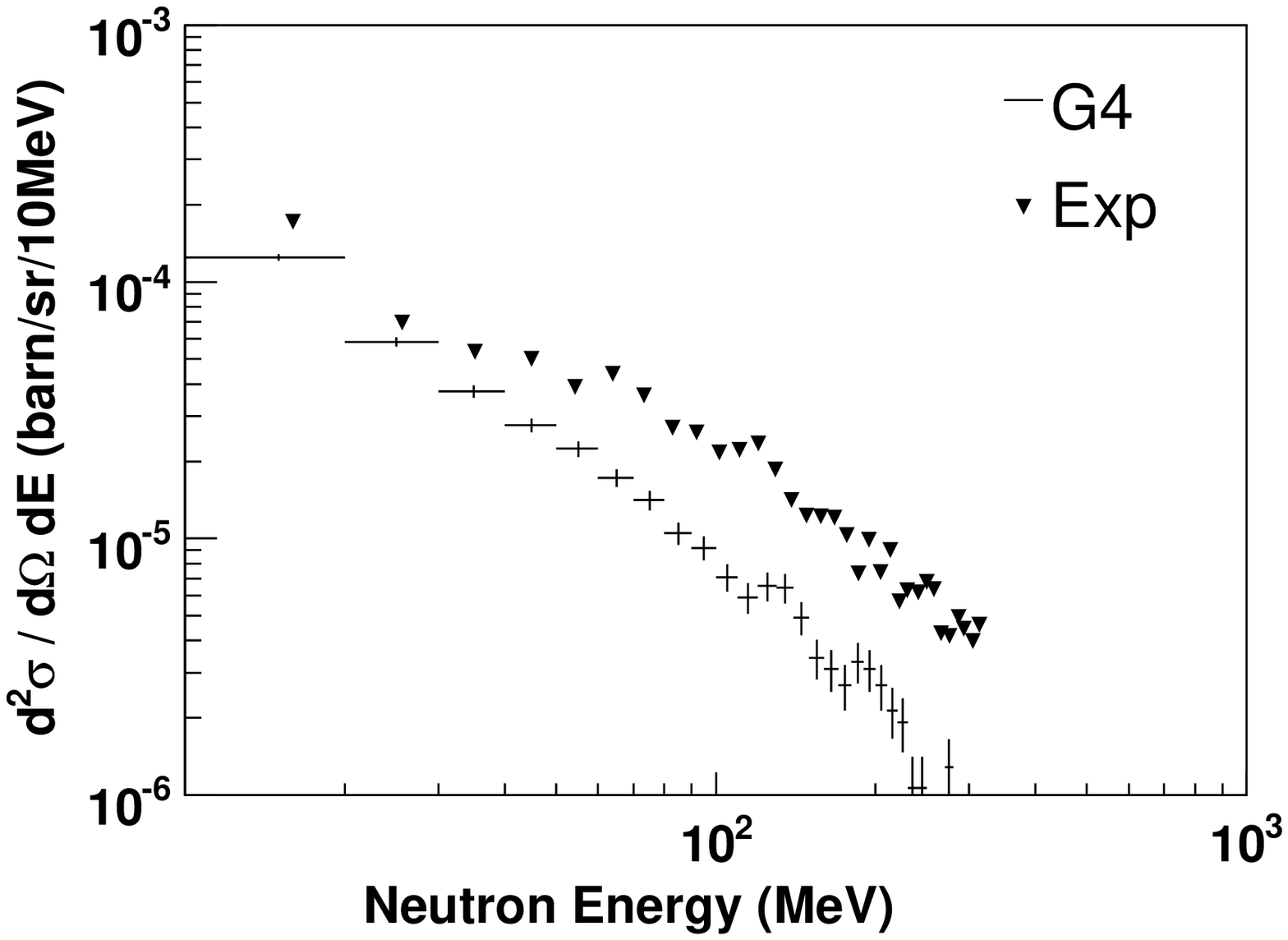}}
	\subfigure[$\theta=$135$^{\circ}$]{\label{fig:Copper_135Deg} \includegraphics[width=.9\columnwidth]{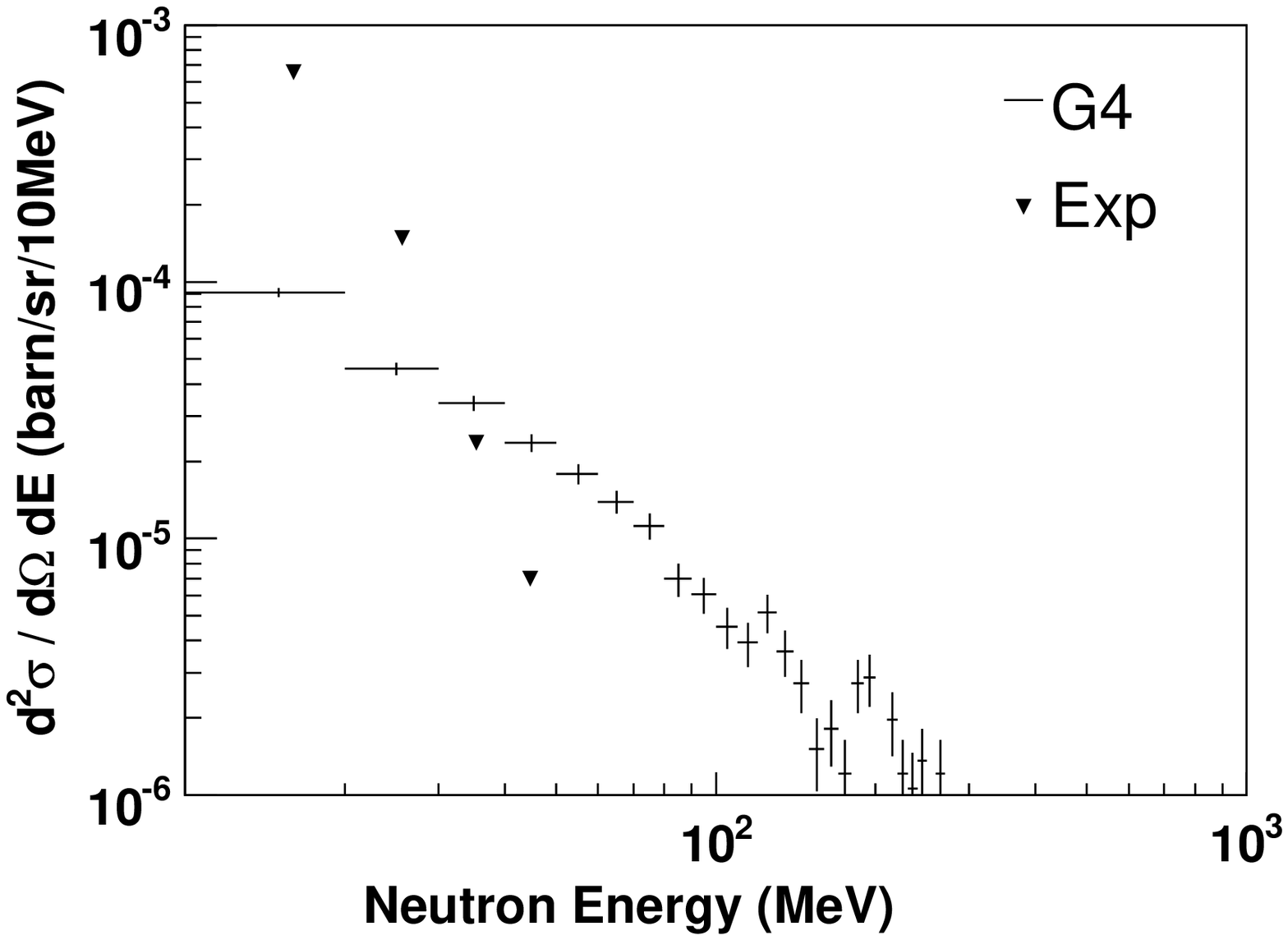}}
	\subfigure[Angular distribution.]{\label{fig:Copper_CosTheta} \includegraphics[width=.9\columnwidth]{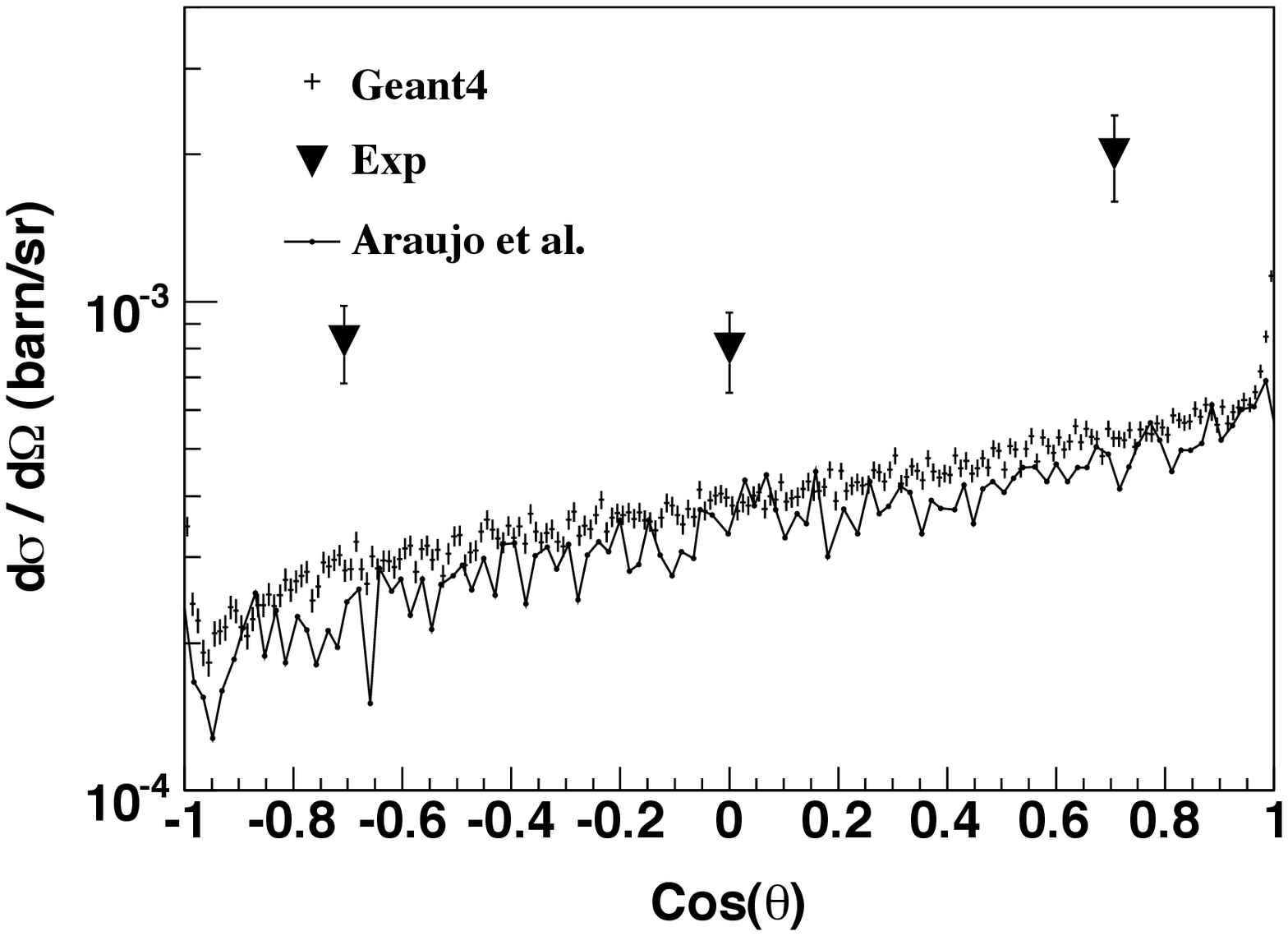}}
\caption{Comparison between the NA55 experiment and simulations for the copper beam dump.  The simulated spectrum at 135 degrees is
significantly harder than the measured spectrum. The curvature in the
angular distribution produced by Geant4 exhibits curvature opposite to observations.
}
\label{fig:CERNCopper}
\end{figure*}

\begin{figure*}[ht]
\centering
	\subfigure[$\theta=$45$^{\circ}$]{\label{fig:Lead_45Deg} \includegraphics[width=.9\columnwidth]{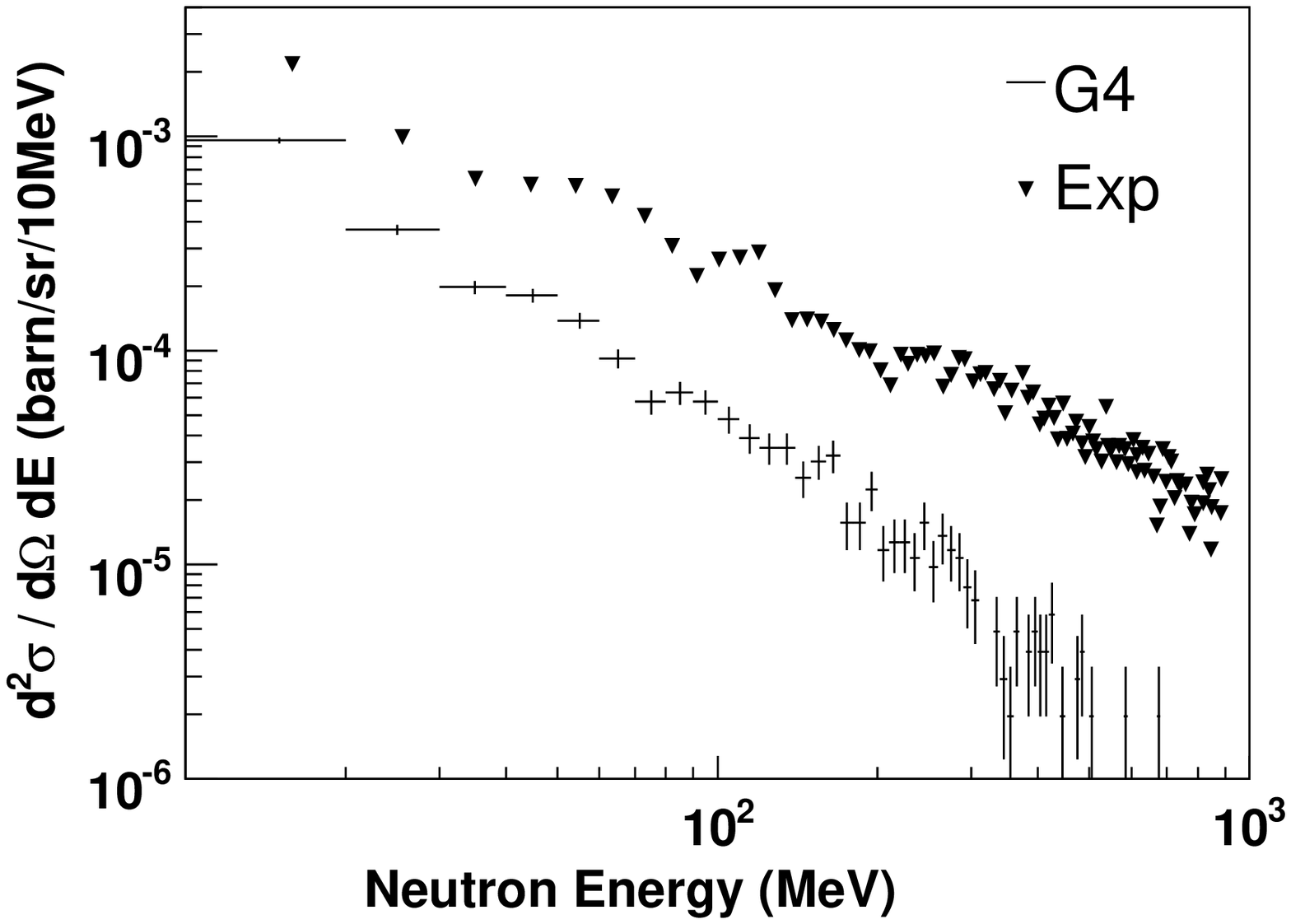}}
	\subfigure[$\theta=$90$^{\circ}$]{\label{fig:Lead_90Deg} \includegraphics[width=.9\columnwidth]{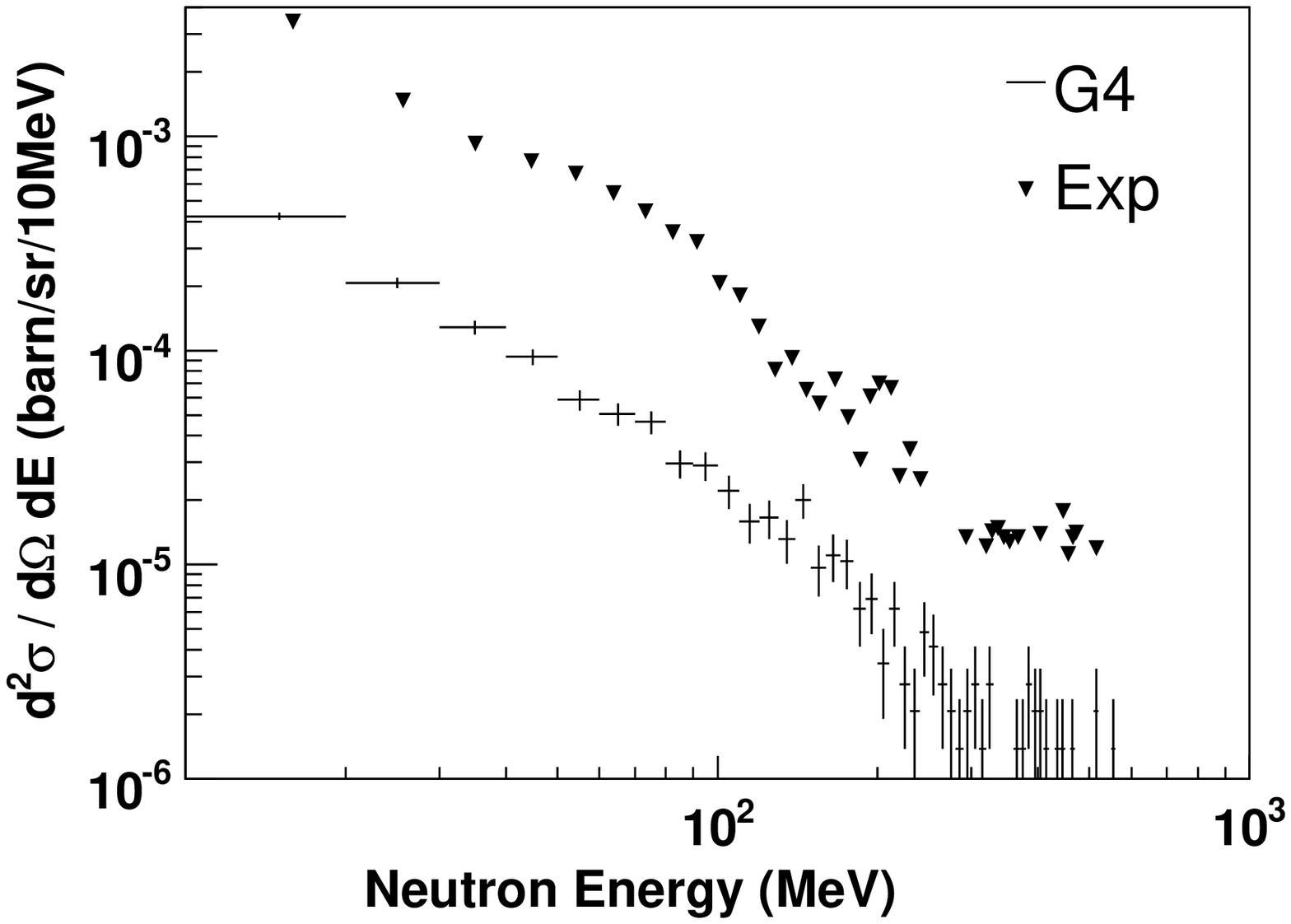}}
	\subfigure[$\theta=$135$^{\circ}$]{\label{fig:Lead_135Deg} \includegraphics[width=.9\columnwidth]{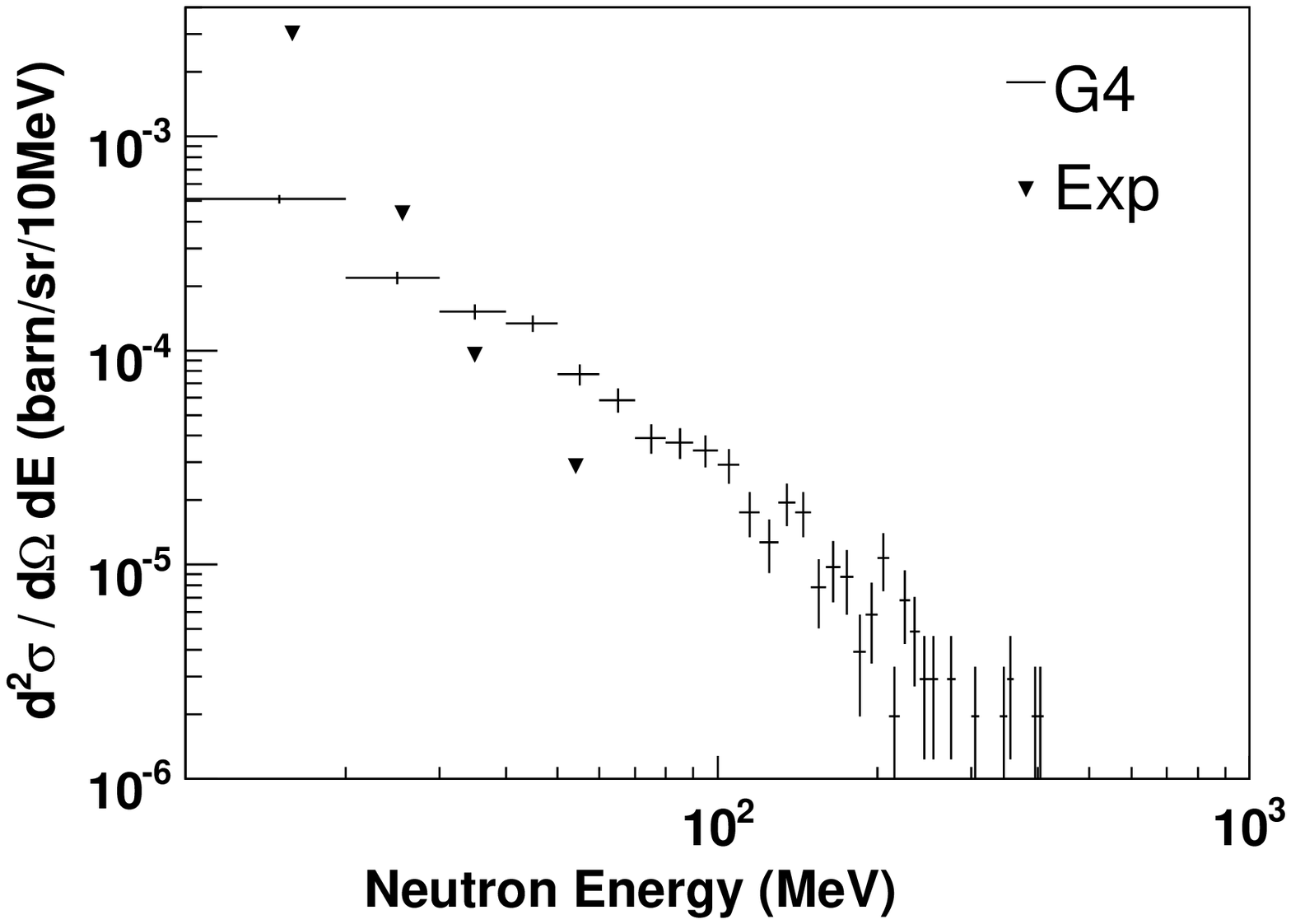}}
	\subfigure[Angular distribution.]{\label{fig:Lead_CosTheta} \includegraphics[width=.9\columnwidth]{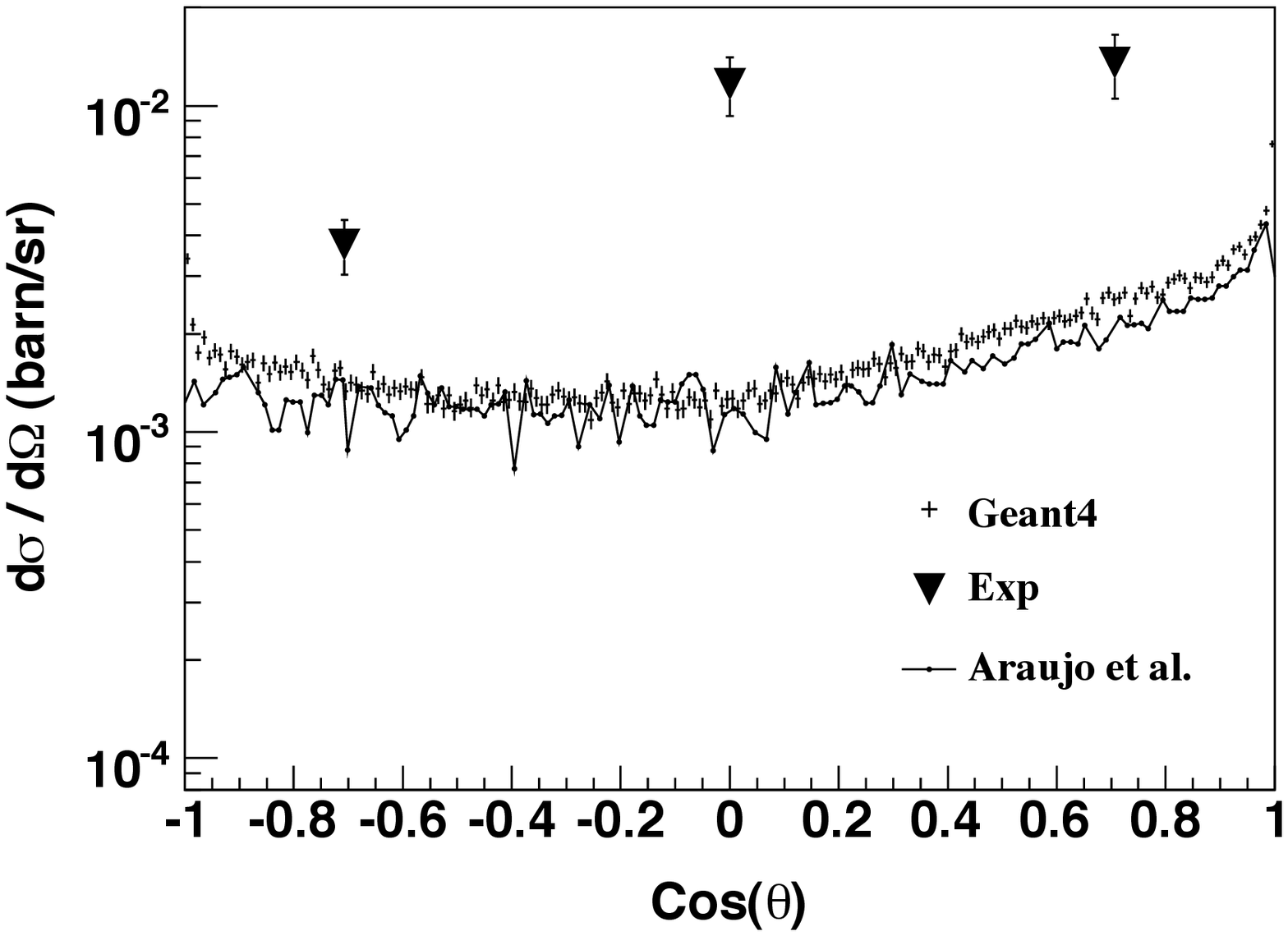}}
\caption{Comparison between the NA55 experiment and simulations for the lead beam dump.  
The agreement for lead is worse than that for graphite and copper (see Figs.~\ref{fig:CERNGraphite},\ref{fig:CERNCopper}).  The simulated spectrum at 135 degrees is
significantly harder than the measured spectrum. The curvature in the
angular distribution is not reproduced by Geant4.
}
\label{fig:CERNLead}
\end{figure*}

Results are presented in Figures~\ref{fig:CERNGraphite}, \ref{fig:CERNCopper}, and \ref{fig:CERNLead}.  The double-differential cross sections ($d^2\sigma/dE d\Omega$) were generated by selecting neutrons with energies greater than 10~MeV (the threshold of the detectors in the experiment) that satisfied $\theta_0 \pm 2^{\circ}$ for $\theta_0$ equal to 45, 90, and 135 degrees as measured from the forward beam direction.   The results were divided by the angular acceptance ($2\pi\Delta\theta$), the total muons on target, and the cross-sectional composition of the material (measured in atoms/cm$^2$).
To generate the angular distribution ($d\sigma/d\Omega$), the number of neutrons above 10~MeV was binned uniformly in $\cos\theta$ and then scaled appropriately.  Our results for the angular dependence
agree well with the results from both Geant4 and FLUKA reported in~\cite{Araujo:2004rv}.

Total simulated neutron production in units of neutrons/muon/g/cm$^2$ can be determined by integrating the results of the generated differential neutron-production cross section.  Determining the total experimental neutron production is not possible given that a very small subset of the solid angle space was
covered.  To estimate this quantity, the measured angular distributions were fit to a straight line which was then integrated.  Results of this and comparison to
simulated numbers are presented in Table~\ref{tab:CERNNeutronProd}.  

Since the NA55 experiment used thin targets in which the hadronic and electromagnetic
showers induced by the muons do not fully develop, it is difficult to compare directly to results presented previously which considered muons through thick
material (see, for example \cite{Wang:2001fq,Wulandari:2004bj,Araujo:2004rv} and references therein).  However, it is possible to  compare corresponding ratios
(e.g.~ratio of Geant4 to experiment) between this work with thin targets and previous results with thick targets.  The comparison of these ratios is important
when trying to determine the source of the discrepancy between simulated results and experimental data.  For example, equivalent ratios between simulated and
experimental data for thin and thick targets would suggest that hadronic showers are handled adequately within the simulation.  
in~\cite{Araujo:2004rv}, it is found that Geant4 is a factor of 2 low with respect to experimental results of muon-induced neutron production within liquid scintillator
(C$_{10}$H$_{20}$) (thick target) which has an effective atomic mass close to that of graphite.  
This ratio agrees well with the ratio calculated in this work (a factor of~2, see Table~\ref{tab:CERNNeutronProd}). 
 For higher-Z materials (e.g.~lead), Wulandari et al.~\cite{Wulandari:2004bj} assumed that the neutron production versus muon energy follows the power law
observed for liquid scintillator ($\sim$E$^{0.75}$) and then extrapolate from experimental data at different energies (110~GeV, 385~GeV) to an energy (270 GeV,
LNGS depth) where simulations have been performed.  With this extrapolation method, they found agreement with data within a factor of~2 for both FLUKA and Geant4
for lead~\cite{Wulandari:2004bj}.  This disagrees with the ratio found in this work (5.9, see Table~\ref{tab:CERNNeutronProd}) which suggests that Geant4
overproduces neutrons in hadronic and electromagnetic showers to compensate for initial underproduction of neutrons in lead.  This conclusion
is also noted by Ara\'{u}jo et al.  

The comparison between simulated and measured double-differential cross sections indicates a growing disagreement as the atomic mass of the target material increases.  In particular, simulated results from graphite (see Figure~\ref{fig:CERNGraphite}) show a reasonably good agreement with the energy spectrum for 45, 90 and 135~degrees though the simulated results for 135~deg produce a harder spectrum.  The agreement worsens for Cu and Pb (see Figures~\ref{fig:CERNCopper},\ref{fig:CERNLead}), which exhibit significantly over-hard spectra at 135~degrees.  For lead and copper, the simulated angular distribution shows opposite curvature to the measurement. This was also noted in~\cite{Araujo:2004rv}.

\begin{table*}
\caption{Experimental and simulated neutron productions for the CERN NA55 experiment.  Experimental values were estimated by integrating a linear fit to differential cross section measurements at 45, 90, and 135~degrees.}
\label{tab:CERNNeutronProd}
\centering
\begin{tabular}{l|c|c|c}
Material & Exp. Fluence [n/$\mu$/g/cm$^2$] & Calc. Fluence [n/$\mu$/g/cm$^2$]  & Ratio \\
\hline
\hline
Graphite & 6.8$\times10^{-5}$ & 3.3$\times10^{-5}$ & 2.1 \\
Copper & 1.2$\times10^{-4}$ & 4.8$\times10^{-5}$ & 2.5 \\
Lead & 3.5$\times10^{-4}$ & 5.9$\times10^{-5}$ & 5.9 \\
\hline
\end{tabular}
\end{table*}

\section{SLAC Experiment}

An experiment was performed at the Stanford Linear Accelerator Center (SLAC)~\cite{SLAC1} which involved a 28.7~GeV electron beam incident upon an aluminum beam
dump.  Primary neutrons were generated mainly by photonuclear interactions initiated by bremsstrahlung photons from the decelerating electrons.  The neutron
time-of-flight and energy spectra were measured outside a steel-and-concrete shield in which the concrete width was varied.
The final neutron spectrum outside the shield consisted of both primary neutrons as well as secondary neutrons created through inelastic interactions of the primary neutrons with shield material.  The experiment was simulated in detail using the FLUKA simulation package~\cite{SLAC2} generating results in agreement with experimental data.  

\begin{figure}[htp]
\centering
\includegraphics[width=.9\columnwidth]{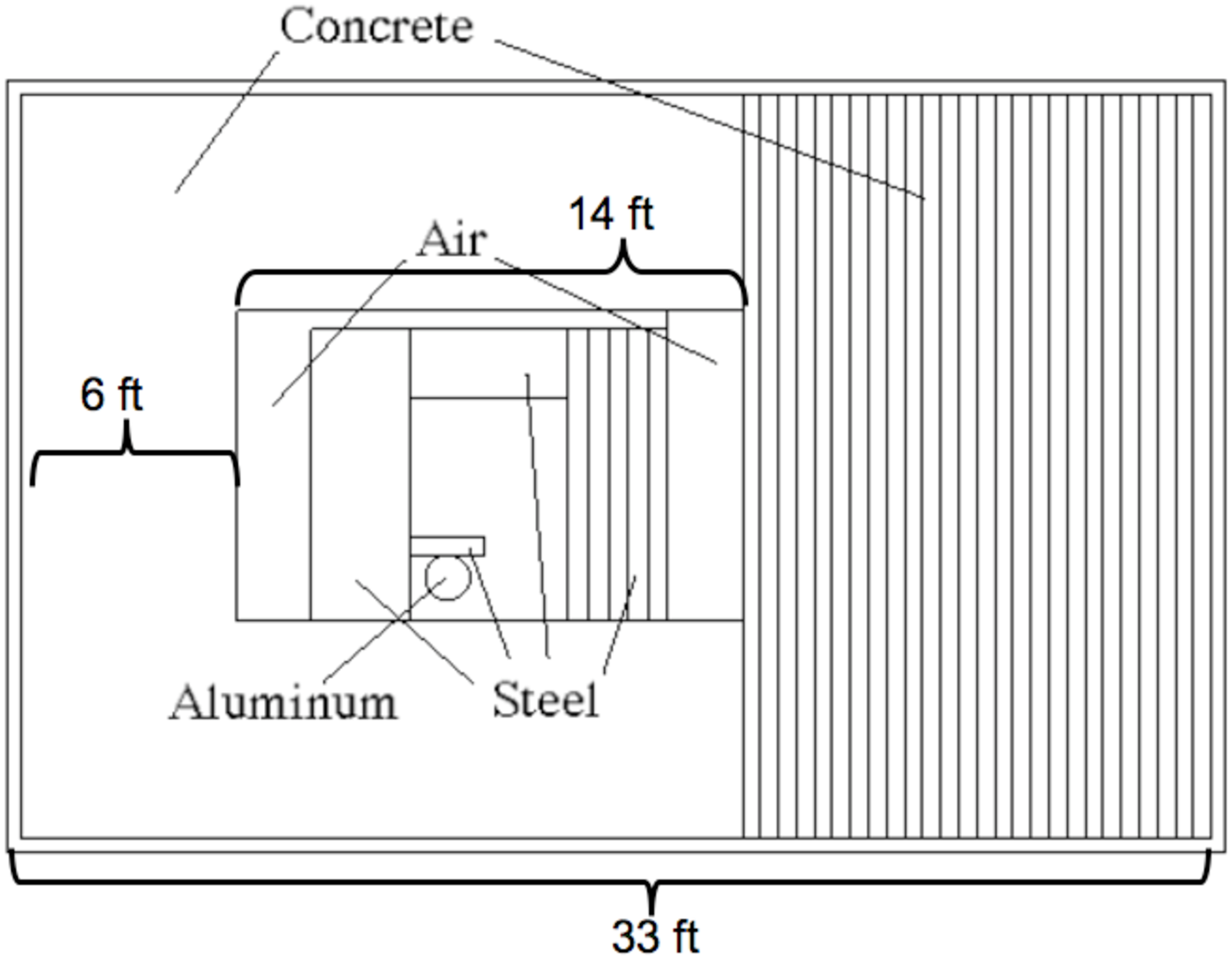}
\caption{A beam view of the simulated geometry.  The vertical lines in the concrete and steel materials indicate regions of different importance which are placed to facilitate the variance reduction technique importance sampling employed in this work.  In this view, the detector is located to the right of the shielding.}
\label{fig:BeamView}
\end{figure}

A detailed description of the SLAC beam dump experiment can be found in~\cite{SLAC1}.  Included are schematic diagrams from~\cite{SLAC1} demonstrating the simple design employed by the experiment (see Figs.~\ref{fig:BeamView}).  A 28.7~GeV electron beam was incident upon a cylindrical aluminum beam dump inside a shield housing with an opening to allow the beam to enter.  The shield housing consisted of two lateral steel shields and one top steel shield in addition to concrete positioned around all sides.  An organic liquid scintillator detector was placed outside the shields laterally from the aluminum beam dump and the neutron energy and time-of-flight (TOF) spectra were measured for shield concrete shield widths of 9~ft, 11~ft, and 13~ft (2.74~m, 3.35~m, and 3.96~m).  

\begin{table*}
\caption{A description of the composition of materials in the SLAC beam dump.  Fractions are given in mass percentage and relative atom number as indicated.
}
\label{tab:MatComp}
\begin{center}
\begin{tabular}{c|c|c|c|c}
\hline
Material & Density g/cm$^3$ & Element & A (g/mol) & Fraction \\
\hline
\hline
\multirow{10}{*}{Concrete} & \multirow{10}{*}{2.35} & O & 16.00 & 50.0 (mass \%) \\
& & Si & 28.06 & 20.0 \\
& & Ca & 40.08 & 19.5 \\
& & Al & 26.98 & 3.0 \\
& & C & 12.01 & 3.0 \\
& & Fe & 55.85 & 1.4 \\
& & Na & 22.99 & 1.0 \\
& & K & 39.10 & 1.0 \\
& & H & 1.01 & 0.6 \\
& & Mg & 24.31 & 0.5 \\
\hline
\hline
\multirow{3}{*}{Steel} & \multirow{3}{*}{7.5} & Fe & 55.85 & .7 (rel. atom \#) \\
& & Ni & 58.69 & .2 \\
& & Cr & 52.00 & .1 \\
\hline
\hline
Aluminum & 2.70 & Al & 26.98 & 100 \% \\
\hline
\end{tabular}
\end{center}
\end{table*}

\begin{table*}
\caption{Effective A and Z for various underground sites as compiled by Ref.~\cite{MEI06}.  A comparison to concrete used in the SLAC experiment demonstrates its similarity in character to rock in underground sites around the world.}\label{tab:MatCompToUnderground}
\begin{tabular*}{0.8\textwidth}{@{\extracolsep{\fill}} lcccc}
\hline
\hline
Site & $\langle A\rangle$ &$\langle Z\rangle$ &$\langle Z\rangle/\langle A\rangle$ & $g/cm^3$\\
\hline
WIPP & 30.0 & 14.64 & 0.488 & 2.3 \\
Soudan & 24.47 & 12.15 & 0.497 & 2.8 \\
Kamioka & 22.0 & 11.0 & 0.5 & 2.65 \\
Boulby & 23.6 & 11.7 & 0.496 & 2.7 \\
Gran Sasso & 22.87 & 11.41 & 0.499 & 2.71 \\
Sudbury & 24.77 & 12.15& 0.491 & 2.894 \\
\hline
SLAC Concrete & 24.13& 12.0  & 0.497 & 2.35 \\
\hline
\hline
\end{tabular*}
\end{table*}

\subsection{Simulation}\label{sec:SLACSim}

A comprehensive description of the modeled geometry is given in~\cite{SLAC2}; the experimental geometry was modeled within Geant4 as closely as possible
following this description.  A complete list of materials used is given in Table~\ref{tab:MatComp}.  A comparison of the composition of the concrete to typical
rock at selected underground labs is given in Table~\ref{tab:MatCompToUnderground}.  

To generate meaningful statistics within the limitations of computation time, it was necessary to employ importance sampling, a well-established variance reduction technique (e.g.~see~\cite{Metropolis:1953am}).  
A set of classes implementing importance sampling is included within the Geant4 framework.  Using this implementation required constructing regions in the lateral shield that increased in importance moving away from the beam dump towards the detector.  A neutron moving from a region of lower importance to one of higher importance is multiplied (i.e.~the track is split into multiple copies) according to the ratio of importance values, and its ``weight" is multiplied by the inverse of this ratio.  Conversely, a neutron moving from higher importance to lower importance is killed according to the importance value ratio.  
This has the advantage of effectively multiplying the neutron spectrum at each half foot of shield width and eliminating neutrons that travelled well away from the shield region.  Geant4 keeps track of the weight of each neutron to allow a comparison of the final simulation results with experimental data.  

Tracking cuts were established to kill particles either below certain energies or in regions deemed sufficiently far from the lateral shield on the detector side so as to have little or no contribution to the final results.  In particular, cuts were established to kill low-energy electromagnetic particles such as photons (5~cm), electrons and positrons (1~cm), and generic charged particles (5~mm).  
Geant4 used the distance cuts to calculate energy cuts for the different materials in the simulation.  For example, these distance cuts corresponded to energy cuts of $\sim$50~keV ($\sim$4~MeV) for gammas (electrons and positrons) for concrete.   Within steel, these cuts became $\sim$200~keV ($\sim$12~MeV) respectively.  In addition, all neutrons with a kinetic energy below 900~keV were removed.  
Geant4 does not normally provide an energy cut for neutrons and so this cut was implemented within the UserSteppingAction method called at each step.  Since simulating the thermalization of neutrons can occupy significant amounts of computation time, this last cut reduced cpu usage.  The 900~keV energy cut on neutrons was well below the experimental neutron energy threshold (6~MeV).  This cut also eliminated gammas produced in neutron-capture reactions, which were of too low energy to contribute to the neutron flux through photonuclear interactions.  Lastly, all particles entering a region 15~cm into the lateral steel shield opposite to the detector were killed, since it was deemed these particles would not contribute significantly to the flux at the detector.  

The simulations were performed for all three shield widths (9~ft, 11~ft, and 13~ft).  As a neutron crossed every half foot in the shield, the energy, absolute
time (from the initial electron event), position, momentum vector, and weight were recorded.   From these parameters, we calculated total fluence in addition to
neutron energy spectra and time-of-flight (TOF) distributions at each shield width.  Results of these calculations are presented below.

\subsection{Energy Spectra}

\begin{figure*}[ht]
\centering
	\subfigure[9 ft.]{\label{fig:Energy_9ft} \includegraphics[width=.9\columnwidth]{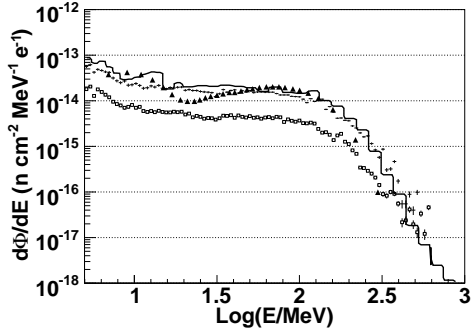}}
	\subfigure[11 ft.]{\label{fig:Energy_11ft} \includegraphics[width=.9\columnwidth]{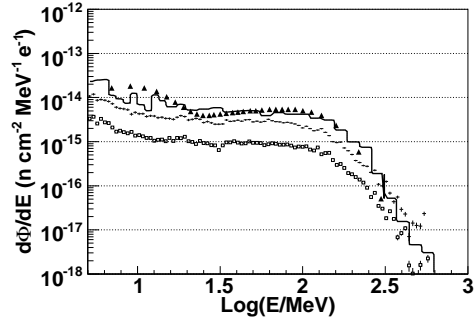}}
	\subfigure[13 ft.]{\label{fig:Energy_13ft} \includegraphics[width=.9\columnwidth]{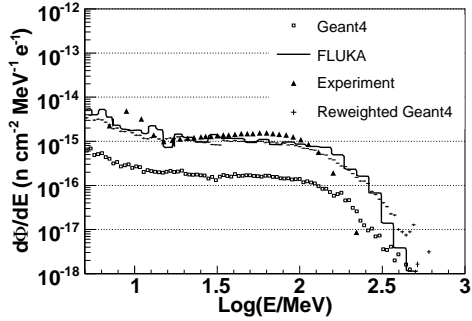}}
\caption{A comparison of calculated and experimental energy spectra for three shield widths. FLUKA and experimental data were obtained from~\cite{SLAC2}.
Statistical error bars are included on both of the Geant4 histograms but not on the spectra from FLUKA and experiment.  An explanation of Reweighted Geant4 is
given in Section~\ref{sec:G4RW}.} 
\label{fig:Energy}
\end{figure*}

The Monte Carlo energy spectra were generated for neutrons at or above 5~MeV, normalized per incident beam electron.  Experimental data were gathered
for neutrons above 6~MeV and the detector response was unfolded from the raw measurements to compare directly with calculated results.  Comparisons between the
Geant4 calculated spectra, the FLUKA calculated spectra, and experimental data are shown in Figure \ref{fig:Energy}.  Geant4 demonstrates reasonably good
agreement with the shape of the spectra, but the agreement with the fluence slightly worsens as the shield width is increased (this will be discussed further in
Section~\ref{sec:SLACFluence}).  Geant4 and FLUKA both calculate harder spectra than the experimental data; this may arise due to a lack of understanding of
detector response at high neutron energy and limited experimental statistics at these energies (see~\cite{SLAC2}).  

\subsection{TOF Distribution}
Calculating the TOF distributions provided a useful consistency check for the simulation.  It was found that the TOF distributions reproduced the shape of the experimental data, but also demonstrated slightly diminishing fluence with increasing shield width (see Section~\ref{sec:SLACFluence}).  This was consistent with results for the energy spectra.

\subsection{Fluence}\label{sec:SLACFluence}

\begin{figure*}[ht]
\centering
\includegraphics[width=.9\textwidth]{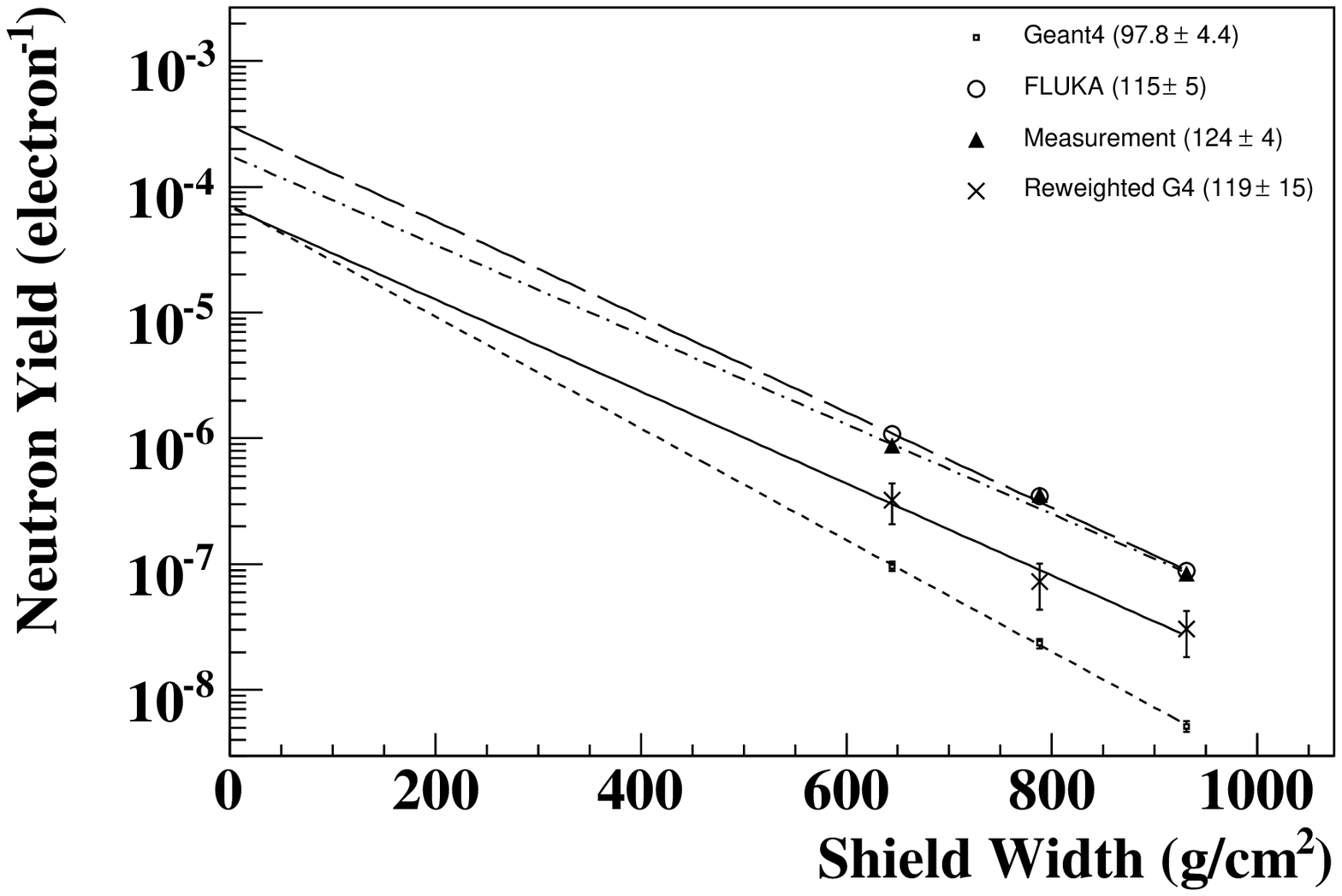}
\caption{A comparison of calculated (Geant4 and FLUKA) and measured total neutron fluence (E$\geq6$~MeV) for varying shield width.  Exponential fits of the form
$A\exp^{-x/\lambda}$ (where $x$ is the length through the shield) to each set of points are drawn and the values of the fit characteristic lengths ($\lambda$) are included in parentheses in units of
g/cm$^{2}$.  See Table~\ref{tab:FluxCompare} for more details.  Reweighted Geant4 is explained in more detail in Section~\ref{sec:G4RW}.  It is important to
note that the technique employed in Reweighted Geant4 only corrected the attenuation length (slope in the above figure) within the concrete and therefore does
not correct for the discrepancy in total fluence between Geant4 and data at zero shield width.} \label{fig:Fluence}
\end{figure*}

The total fluence of neutrons having kinetic energy 6 MeV and greater was calculated for each of the three shield widths and compared to the values from FLUKA and experiment (see Figure~\ref{fig:Fluence} and Table~\ref{tab:FluxCompare}).  Exponential fits were performed for all three data sets.  These fits show that the attenuation length of neutrons simulated by Geant4 is shorter than measured in the experiment.  The fits also indicate that the total neutron flux entering the concrete shield (i.e.~at a shield width of 0~ft) is underestimated by Geant4 by about a factor of 4-5.  Neutrons entering the concrete also pass through a steel shield of about a meter in width.  Since no data were taken with only the steel shield, a direct comparison between simulated and experimental neutron production is not possible.

\begin{table*}
\caption{Comparison of fluence values for various shield widths.  FLUKA and measurement results were obtained from Ref.~\cite{SLAC1}.  The $r$ value corresponds to
the distance from the detector to the center of the beam dump.  The square of this value is multiplied by the fluence to obtain the yield.  Errors on the Geant4
fluence are statistical. 
The yields for each set of results were then fit to an exponential using chi-square minimization to obtain the attenuation length.  The results of these fits are
discussed in more detail in section~\ref{sec:G4RW}.  Reference~\cite{Jenkins:1979} cites the value of $\lambda$ as being 120~g/cm$^2$ for concrete.  Reweighted
(RW) Geant4 is a technique used to help correct for the discrepancy seen in the attenuation lengths and it is described in more detail in Section~\ref{sec:G4RW}.
The fit for RW Geant4 is further constrained by demanding that the fluences of Geant4 and RW Geant4 match (within errors) at zero concrete shield
width.}\label{tab:FluxCompare} 
\begin{centering}
\begin{tabular*}{0.9\textwidth}{@{\extracolsep{\fill}}lccccc}
\hline
\hline
Source & Shield Width [cm] & Fluence [e$^{-1}$ cm$^{-2}$] & $r$ [cm] & Neutron Yield [e$^{-1}$] & $\lambda$ [g/cm$^2$] \\
\hline
\hline
FLUKA & 274 & $3.59 \times 10^{-12}$ & 549 & $1.08 \times 10^{-6}$ \\
& 335 & $9.30 \times 10^{-13}$ & 610 & $3.46 \times 10^{-7}$ & $115 \pm 5$ \\
& 396 & $1.97 \times 10^{-13}$ & 671 & $8.87 \times 10^{-8}$ \\
\hline
Measured & 274 & $2.91 \times 10^{-12}$ & 549 & $8.77 \times 10^{-7}$ \\
& 335 & $9.62 \times 10^{-13}$ & 610 & $3.58 \times 10^{-7}$ & $124 \pm 4$\\
& 396 & $1.88 \times 10^{-13}$ & 671 & $8.46 \times 10^{-8}$ \\
\hline
Geant4 & 274 & $3.21 \pm 0.29 \times 10^{-13}$ & 549 & $9.67 \pm 0.88 \times 10^{-8}$ \\
& 335 & $6.32 \pm 0.54 \times 10^{-14}$ & 610 & $2.35 \pm 0.20 \times 10^{-8}$ & $97.8 \pm 4.4$\\
& 396 & $1.14 \pm 0.11 \times 10^{-14}$ & 671 & $5.15 \pm 0.50 \times 10^{-9}$ \\
\hline
Reweighted Geant4 & 274 & $1.07 \pm  0.38 \times 10^{-12}$ & 549 & $3.23 \pm 1.16 \times 10^{-7}$ \\
& 335 & $1.95 \pm 0.77 \times 10^{-13}$ & 610 & $7.24 \pm 2.89 \times 10^{-8}$ & $119 \pm 15$\\
& 396 & $6.74 \pm 2.67 \times 10^{-14}$ & 671 & $3.03 \pm 1.20 \times 10^{-8}$ \\
\hline
\hline
\end{tabular*}
\end{centering}
\end{table*}

\section{Estimating systematic errors of the simulations}

Given the inconsistencies found in the comparison of neutron results from Geant4 and experiment, it is important to quantify systematic errors and, if possible, provide a method for correcting for these discrepancies.  It should be emphasized that the techniques presented in this section are not suggestions for permanent fixes, but rather methods that may be used cautiously until as the Geant4 physics models are improved to better match observations.

\subsection{Muon-induced neutron production}

A few conclusions may be drawn from the
results relevant to simulations attempting to estimate muon-induced
backgrounds with Geant4. For forward-scattered neutrons, the shapes
of the energy and angular distributions are reasonably well
reproduced, albeit with a production cross section too low by a factor
of $\sim$2 in low-Z material.   For lead, this cross section is too low by a factor of 6.  
Until the source of this discrepancy in the physics models can be
corrected, it will be necessary to correct for the neutron deficit by
applying a multiplicative factor to the normalization. The uncertainty of this correction should be taken to be about on the order of the size of the correction itself.  Note that Mei
and Hime~\cite{MEI06} found it necessary to apply a similar
muon-energy-dependent neutron-multiplicity correction that corresponds
to a factor of 1.5 at 190~GeV in order to correct for an apparent
too-low spallation neutron yield for thick liquid scintillator targets
in FLUKA.

    The situation for backscattered neutrons in Geant4 is much more difficult to assess, since for higher Z the simulated spectrum is significantly harder than
the experimental spectrum. On one hand, a harder spectrum can be expected to induce more interactions with a simulated detector, in which case one might consider
such a background estimate to be more ``conservative."  On the other hand, for detectors that are sensitive to higher multiplicity events, such increased activity
could artificially enhance the efficiency for vetoing such backgrounds, leading to an underestimation of background. Hence we urge extreme caution when simulations are
sensitive to backscattered neutrons, for example in simulations of ground-shine backgrounds.

\subsection{Neutron propagation}\label{sec:G4RW}

Simulation of neutron transport through a low-Z material (concrete) indicates a diminishing agreement with experiment as neutrons traverse a larger amount of
material.  Therefore, an estimation of systematic errors in neutron transport must take into account the dimensions of the material present within the
simulation.  Because the shape of the neutron energy spectrum is maintained, albeit with a slightly harder spectrum, it is reasonable to estimate the systematic
error using the integrated neutron fluxes listed in Table~\ref{tab:FluxCompare}.  For example, for a material of 10~ft width, Geant4 over-attenuates neutrons by
roughly a factor of 4.  Estimating this error for a particular experiment would involve some determination of average length traversed by neutrons, but this
could prove difficult especially for those experiments involving non-trivial geometries.

To the authors' best knowledge, no data is available to which one may compare a simulated neutron transport through high-Z material (e.g.~lead).  Some analyses of the cascade models important for such a simulation have
been performed~\cite{Ivanchenko:2003wn}, but these studies have only analyzed incident protons.
Therefore, to estimate neutron transport errors for high-Z material one must extrapolate from results at low-Z.

To account for the over-attenuation of neutrons in low-Z material in Geant4, it is possible to ``reweight" the neutrons as they traverse the concrete.  One should note that this technique strays from the interaction model used within the Geant4 framework since the reweighting is based upon patching the demonstrated discrepancy between results of Geant4 and data.  Despite this, the authors believe this technique provides a method by which one may more confidently estimate background arising from neutrons which must propagate through significant amounts of low-Z material until the source of the
over-attenuation is addressed in the underlying physics models.

The simple technique takes advantage of Geant4's track weighting.  For each step a neutron takes
through the concrete, the step length is used to calculate a reweighting factor, $R$:
\[
R = e^{-\alpha x}
\]
where $\alpha$ is a constant with units of inverse length and $x$ is the step length of the neutron.  $\alpha$ is calculated using exponential fits to data and simulation results and is equal to the difference between the inverse attenuation lengths in each fit, i.e.~$\alpha = 1/\lambda_{exp} - 1/\lambda_{G4}$ is negative when the simulation over-attenuates neutrons.  The weight of the track is multiplied by $R$ (which is greater than~1) at each step.  The weight of the neutron is propagated to any secondary neutrons produced in subsequent interactions.  This ensures that any simulation attempting to estimate neutron-induced background will appropriately count the background from both primary and secondary neutron interactions.  

Results of the reweighting technique are presented within Figs.~\ref{fig:Energy} and~\ref{fig:Fluence} and Table~\ref{tab:FluxCompare}.  Here,
a value of $\alpha = -0.48 \pm 0.09$~m$^{-1}$ was used.  The error on alpha arises from the errors on $\lambda_{exp}$ and $\lambda_{G4}$ as quoted in
Table~\ref{tab:FluxCompare}.  Reweighting introduces an additional variance to the neutron flux since the technique depends upon the integrated path length of each neutron.  This additional variance was estimated by analyzing the distribution of neutron path lengths and included in the error bars on the reweighted total flux.  To minimize the effect of this on the fit of the fluence, it was additionally required that the fluences from both Geant4 simulations match
within errors at the initial concrete shield boundary.  This assumption is justified by the fact that no reweighting was performed outside the concrete shield.
The matching was performed using the fit to the (non-Reweighted) Geant4 results to estimate the neutron yield at zero shield width.  This point was then added to
the three calculated points in the Reweighted Geant4 results, and the set of four points was fit to an exponential function.  The fit for Geant4 (RW
Geant4) resulted in a $\chi^2$ value of $0.24$ ($0.43$) which  for 1 (2) degree(s) of freedom yields a P-value of 0.62 (0.81).  The fit value of $\lambda$ for RW Geant4 is consistent with the measured value within errors.  

The figures in Fig.~\ref{fig:Energy} demonstrate that the reweighting technique effectively reduces the attenuation of neutrons while
maintaining the shape of the energy spectrum.  However, this comes at a cost of increased variance of the overall calculated flux (see Figure~\ref{fig:Fluence}) which could be partially compensated by increased statistics.  It is important to emphasize that this technique could only correct for the attenuation of neutrons in concrete
and not for the overall fluence disagreement seen at zero concrete shield width.  Since no other measurements were made of the initial neutron production within the aluminum or neutron attenuation within the steel shield, it is not possible to ascertain the source of this discrepancy.  However, given the underproduction of neutrons generated from muons (see Section~\ref{sec:CERN}) and the over-attenuation of neutrons in concrete, it is likely that both contribute to the observed difference.   

Since the reweighting technique relies upon a comparison of data and simulated results, it is limited in application to those materials for which neutron
propagation data exists.  The presented results from concrete may be applied to
simulated rock given that the rock has a comparable composition to that of the simulated concrete (see Tables~\ref{tab:MatComp}
and~\ref{tab:MatCompToUnderground}).

\section{Conclusion}

Geant4 simulations involving neutrons have been performed and compared to experimental data.  Systematic errors within the hadronic physics models in
Geant4 have been estimated.  It has been found that Geant4 underproduces neutrons from 190~GeV muons by a factor of $\sim2$ for low-Z material.   The
underproduction is more pronounced in higher-Z material.   The shape of the muon-induced neutron angular energy spectra generated by Geant4 reasonably agrees
with the measured shape for low-Z.   This suggests that the underproduction may be corrected by applying a multiplicative factor.  For higher-Z materials like lead, the neutron underproduction by Geant4 is more pronounced, and the simulated backscattered neutron energy
distribution is much harder than is observed. More work will be required to correct the neutron production in high-Z materials.

Geant4 transports neutrons adequately though low-Z material, though with a slightly larger attenuation than is seen in experiment.  This could have a significant
effect on simulations designed to estimate neutron backgrounds within underground experiments.   It is possible to correct for this over-attenuation by applying a
reweighting technique.  It should be stressed that this correction is intended to be only a temporary fix until the source of the problems in the underlying physics models in Geant4 can be addressed.   

\section{Acknowledgments}

The authors would like to thank the MaGe group of the GERDA and \textsc{Majorana} collaborations for insightful comments and suggestions.  This work was sponsored in
part by the US Department of Energy under Grant Nos.~DE-FG02-97ER41020 and DE-AC02-05CH11231.  This research used the Parallel Distributed Systems Facility at the National Energy Research Scientific Computing Center, which is supported by the Office of Science of the U.S. Department of Energy under Contract No.~DE-AC02-05CH11231.

\bibliography{Neutrons}
\bibliographystyle{h-elsevier}

\end{document}